# A Compact Hybrid Battery Thermal Management System for Enhanced Cooling

Zhipeng Lyu[a+], Jinrong Su[b+], Zhe Li[a+], Xiang Li[a], Hanghang Yan[b], Lei Chen[b]

[a] *School of Mechanical Engineering, Yangtze University, Jingzhou, China*
[b] *Department of Mechanical Engineering, University of Michigan, Dearborn, USA*

+ These authors contributed equally to this work.

* Corresponding author: leichn@umich.edu

**Abstract**

Hybrid battery thermal management systems (HBTMS) combining active liquid cooling and passive phase change materials (PCM) cooling have shown a potential for the thermal management of lithium-ion batteries. However, the fill volume of coolant and PCM in hybrid cooling systems is limited by the size and weight of the HBTMS at high charge/discharge rates. These limitations result in reduced convective heat transfer from the coolant during discharge. The liquefaction rate of PCM is accelerated and the passive cooling effect is reduced. In this paper, we propose a compact hybrid cooling system with multi-inlet U-shaped microchannels for which the gap between channels is embedded by PCM/aluminum foam for compactness. Nanofluid cooling (NC) technology with better thermal conductivity is used. A pulsed flow function is further developed for enhanced cooling (EC) with reduced power consumption. An experimentally validated thermal-fluid dynamics model is developed to optimize operating conditions including coolant type, cooling direction, channel height, inlet flow rate, and cooling scheme. The results show that the hybrid cooling solution of NC+PCM+EC adopted by HBTMS further reduces the maximum temperature of the Li-ion battery by 3.44°C under a discharge rate of 1C at room temperature of 25°C with only a 5% increase in power consumption, compared to the conventional liquid cooling method for electric vehicles (EV). The average number of battery charges has increased by about 6 to 15 percent. The results of this study can help improve the range as well as driving safety of new energy EV.

**Keywords**：Hybrid battery thermal management system**,** Nanofluid coupled phase change material cooling, Pulse flow function, Pumping power consumption

**Highlights**

1. A compact HBTMS integrating U-shaped microchannels and PCM/aluminum foam is proposed.

2. A nanofluid cooling (NC) technology with better thermal conductivity is used

3. A pulse flow function for battery enhanced cooling (EC) is developed.

4. The new NC+PCM+EC hybrid cooling technology can increase the number of battery charges by about 6% to 15% compared to traditional liquid cooling systems.

# 1. Introduction

Rechargeable batteries are the main power energy source for new energy electric vehicles (EV). Among the various types of electrochemical batteries, lithium-ion batteries are widely used due to their high power, high capacity and energy density as well as low self-discharge rate [11-13]. It is worth noting that lithium-ion batteries generate a considerable amount of heat during charging and discharging. EV battery packs (with a large number of cells) are discharged at high rates. Without proper thermal management, high temperatures can damage the battery, resulting in reduced capacity, power, life and safety [14-17]. Conversely, temperatures below the recommended operating range can also affect battery quality and driving safety [18,19]. In order to make the power battery inside the new energy electric vehicle work stably, a solution called battery thermal management system (BTMS) has been proposed. It plays a crucial role in the temperature regulation of lithium-ion batteries [20,21]. BTMS have been categorized into three basic types: active cooling systems, passive cooling systems and hybrid cooling systems [22].

Active cooling systems need to provide additional energy to drive different mechanical devices (fans, pumps, etc.). The system then transports the cooling medium to the vicinity of the heat source with the help of these mechanical devices [23]. Typical examples include air cooling [24-26] and liquid cooling [27-29]. Among them, liquid cooling is the most used active cooling method. This is because water has better heat dissipation properties and is economical. So most liquid cooling systems use water as coolant. For example, Luo, *et.al.* [30] designed a center-dispersed square spiral ring (SSR) microchannel liquid-cooling plate for power batteries, to mitigate the temperature gradient effect caused by high localized battery temperatures. The results showed the battery maximum temperature under the thermal management of SSR channel are reduced compared to the existing serpentine channel. Bao, *et.al.* [31] investigated ultra-thin wide DC channel cold plates (WCP) and compared them with serpentine, bifurcated, and U-channel cold plates. The results showed that the temperature standard deviation of WCP was reduced by a maximum of 47.51 %, 45.40%, and 65.08%.

However, because of recent advances in nanomaterials technology, a new heat transfer liquid cooling technology called nanofluids has been developed and applied. Nanofluids are created by dispersing 10 to 50 nm nanoparticles in a classical heat transfer fluid [31]. Many researchers have done analytical studies related to enhanced heat transfer using nanofluids both experimentally and theoretically [32-34]. Compared to conventional fluids, nanofluids exhibit excellent thermal properties. For example, Liu, *et.al.* [35] conducted nanofluid liquid cooling experiments for power batteries to analyze the cooling performance of different nanofluid concentrations and inlet flow rates at different discharge rates. The results show that the cooling performance of the nanofluid is better at low flow rates. At a high flow rate, the cooling ability is poor. This is because high flow rates and high temperatures tend to cause nanofluids to fail, particularly for the high concentration of nanoparticles. Husam, *et.al.* [36] used liquid cooling methods with a variety of nanoparticles ($Al_2O_3$, CuO, $SiO_2$, and ZnO) to reduce the battery temperature. The results show that a higher Reynolds number slightly improves heat transfer at the expense of pumping power. The best cooling effect is obtained by the nanofluid of $SiO_2$ when the Reynolds number value is 18000.

In contrast, passive cooling systems do not require an additional supply of energy. Passive cooling systems include PCM [37-40] and heat pipe cooling [41-43]. The heat dissipation capability of PCM, the primary passive cooling method, is affected by material properties and fill volume. Since the way PCM consumes heat is based on latent heat storage. During phase changes such as melting or solidification, PCM can store or release large amounts of energy in the form of latent heat. In lithium-ion BTMS, the ability of PCM to sustain cooling is reduced when it absorbs heat and melts. This leads to thermal runaway problems such as increased battery surface temperatures and excessive localized temperature differences.

In order to reduce the effects of unfavorable factors in active and passive cooling, the researchers combined the two methods (hybrid cooling system) to achieve greater heat dissipation and higher energy efficiency. For example, Yang, *et.al.* [23] proposed a liquid-cooled plate containing a composite PCM with Z-shaped parallel channels. And a novel delayed cooling strategy was incorporated to investigate a more energy efficient battery thermal management system. The results showed that the optimal hybrid cold plate was designed to weigh only half as much as the baseline cold plate. Total pumping power can be reduced by more than 50% while achieving the same cooling performance. Liu, *et.al.* [44] designed a hybrid battery thermal management system with a honeycomb structure. They analyzed the optimal thermal management performance parameters using liquid flow rate and inlet temperature as variables. The results show that when the coolant flow rate is 0.06 m/s and the inlet temperature is 36°C, the maximum temperature of the battery is 42.3°C. Hybrid BTMS at this condition has the best thermal management performance. Overall, the hybrid cooling system combines the advantages of liquid and phase change cooling with better heat dissipation and energy storage capabilities. And it maintains reliable thermal performance without compromising battery efficiency.

The hybrid liquid cooling system combining active liquid cooling and passive PCM cooling provides an effective solution to solve the thermal management problem of lithium-ion batteries. However, when we review various studies of hybrid cooling systems for battery thermal management, we can see that the fill volume of coolant and PCM in hybrid cooling systems is limited by the size and weight of the BTMS at high discharge rates. These limitations result in reduced convective heat transfer from the coolant during discharge. And the liquefaction rate of PCM is accelerated and the passive cooling time is shortened. From the practical application point of view, the design of a compact hybrid cooling system is desirable for EVs with long range and high performance.

In this paper, we propose a HBTMS integrating multi-inlet U-shaped microchannels and PCM/aluminum foam. To achieve the compactness, the gap between channels is embedded by PCM/aluminum foam. The staggered microchannel/PCM layers are included between rows of battery cells, thus, the battery module is free from cooling plates. Nanofluid cooling (NC) technology with better thermal conductivity is used. A pulsed flow function is further developed for enhanced cooling (EC) with reduced power consumption. A thermal-fluid dynamics model is developed and validated by experiments, which is then used to optimize operating conditions including coolant type, cooling direction, channel height, inlet flow rate, and cooling scheme. The results show that the maximum temperature on the surface of Li-HBTMS is reduced to 38.87°C on average under the novel enhanced method of NC+PCM+EC. This represents a further reduction of 3.44°C compared to the conventional liquid cooling method, which results in a maximum surface temperature of 42.31°C. Furthermore, the pumping power consumption of the system increases by only 5%. Based on these findings, the heat dissipation performance of different cooling schemes is converted into the number of battery charges. The new NC+PCM+EC hybrid cooling method can increase the number of battery charges by about 6% to 15% compared with conventional liquid cooling method.

## 2. Methodology
### 2.1 Physical model

The traditional HBTMS design and the schematic diagram of an integrated U-shaped composite channel for the proposed compact HBTMS are illustrated in Fig. 1. The entire on-board power battery pack consists of several thermal management subsystems in an orderly fashion. A single HBTMS in the wireframe contains 36 18650 cylindrical lithium-ion batteries. The hybrid cooling method consists of a 5-layer composite U-shaped liquid-cooling channels for which the gap between channels is embedded by paraffin phase change material coupled with aluminum foam for compactness. The staggered microchannel/PCM layers are included between rows of battery cells, thus, it is free from cooling plates in the battery module.

Fig. 2 shows a front view and cross-section of a single Li-ion HBTMS structure. The battery cell dimensions are 68 mm × 18 mm (height and diameter). Each 18650 battery is installed in a cylindrical hole in the aluminum of the housing, which has a wall thickness of 1 mm. The hybrid system has liquid cooling channels, and the gap between channels is embedded with PCM/aluminum foam. The height of the cooling channel and phase change are $D$ and $D_1$ ($D = D_1 = 7$ mm), respectively, and the width is 2 mm.

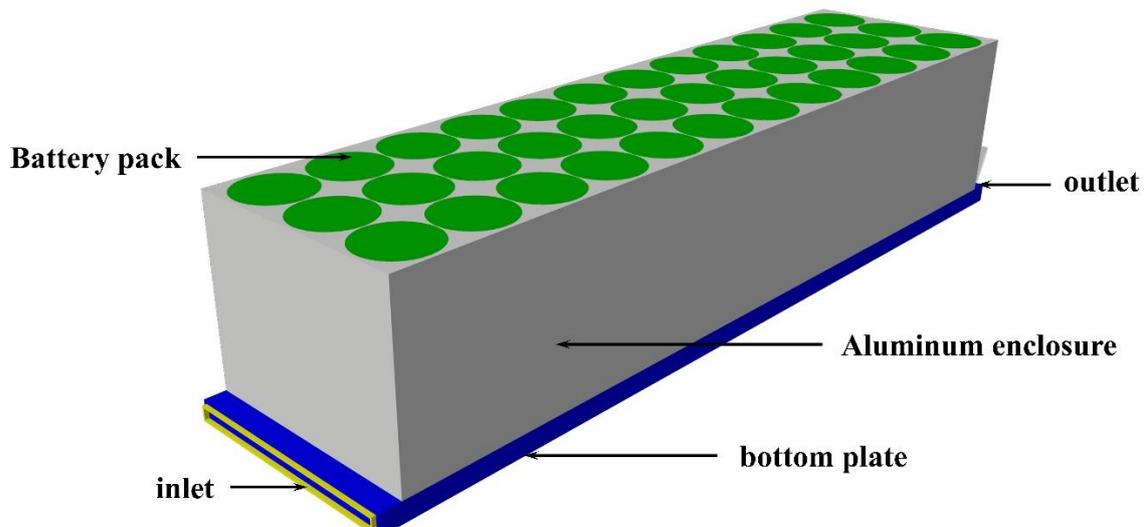

(a) The traditional design

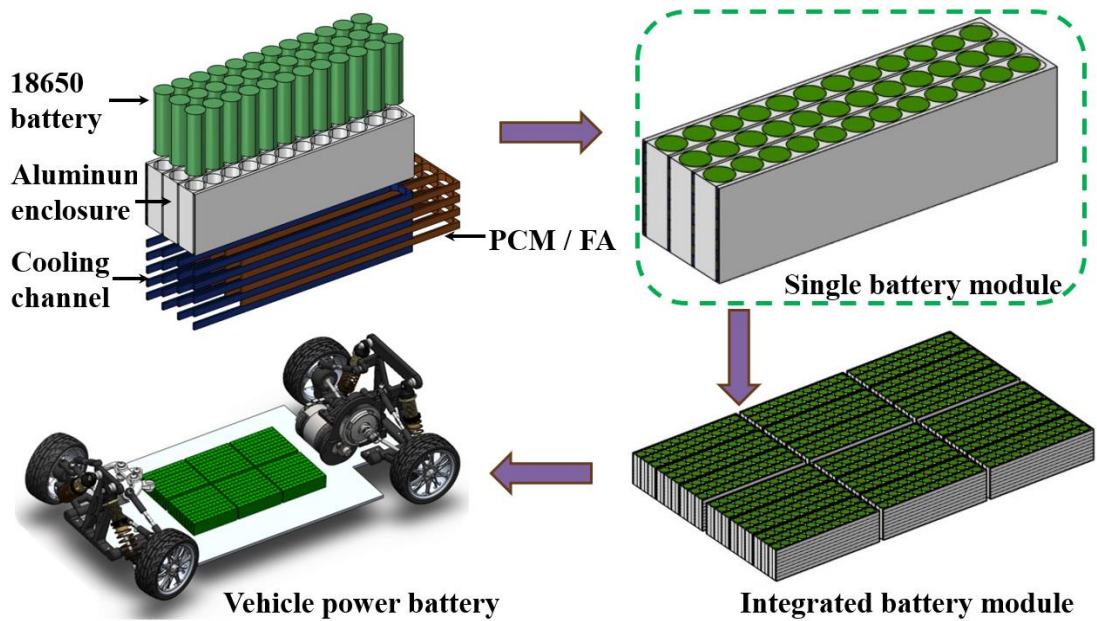

(b) The proposed compact design

**Fig. 1.** Integrated on-board lithium-ion HBTMS

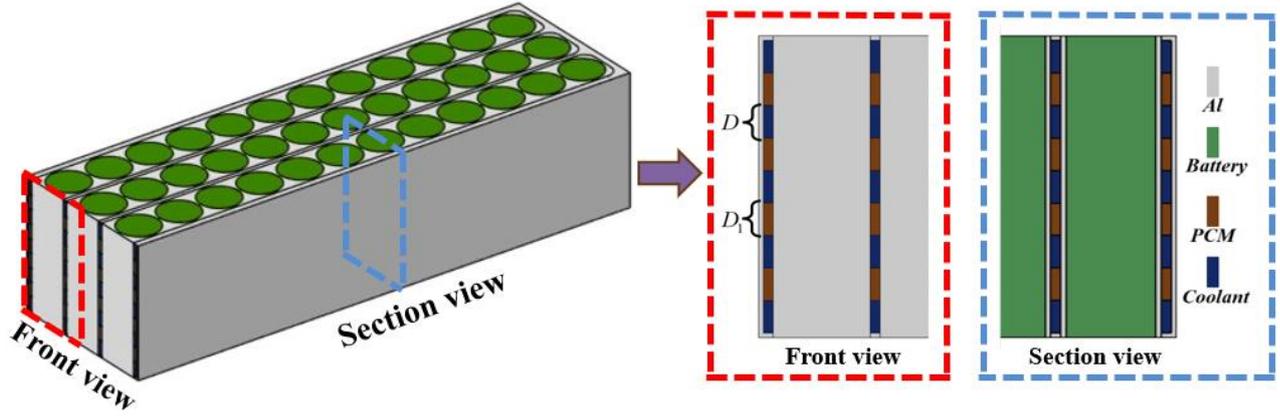

**Fig. 2.** Front view and cross section of HBTMS

## 2.2 Battery heat production model

The energy equation for heat transfer from a lithium-ion battery is as follows:

$$\rho_b c_b \frac{\partial T}{\partial t} = \nabla(k_b \nabla T) + Q_{gen} \qquad (1)$$

where the subscript $b$ denotes the battery. is the battery density. $c_b$ is the specific heat capacity of the battery. $k_b$ is the thermal conductivity of the battery. $T$ is the temperature. $Q_{gen}$ is the rate of heat generation per unit volume inside the battery.

The battery cell is assumed to be a simple homogeneous body with a uniform internal heat generation rate. In general, the battery heat generation rate $Q_{gen}$ can be divided into irreversible heat generated $Q_{ir}$ by the Joule resistance of the battery cell and reversible heat $Q_{re}$ generated by the electrochemical reaction inside the battery. The heat production per unit volume of lithium-ion battery is as follow:

$$Q_{gen} = Q_{ir} + Q_{re} = I^2 R - IT\frac{dE}{dT} \qquad (2)$$

Where $I$ is the charging or discharging current, $R$ is the internal resistance, and $dE/dT$ represents the entropy coefficient depending on the battery state of charge (SOC).

## 2.3 Liquid cooling method

The heat and mass transfer for liquid cooling under this system can be expressed by the three-dimensional continuity, momentum, and energy conservation equations, respectively:

$$\frac{\partial \rho_l}{\partial t} + \nabla \cdot (\rho_l \vec{v}_l) = 0 \qquad (3)$$

$$\rho_l \left[\frac{\partial(\vec{v}_l)}{\partial t} + \vec{v}_l \cdot \nabla \cdot \vec{v}_l\right] = -\nabla p + \mu_l \nabla^2 \cdot \vec{v}_l \qquad (4)$$

$$\rho_l c_l \frac{\partial T_l}{\partial t} + \rho_l c_l \nabla \cdot (\vec{v}_l T_l) = \nabla \cdot (k_l \nabla T_l) \qquad (5)$$

where $\rho, \mu, c$ and $k$ are the density, dynamic viscosity, specific heat capacity and thermal conductivity, separately. $\vec{v}$ is the velocity vector of the coolant, $p$ is the pressure, $T$ is the temperature, $t$ is the time and the subscript $l$ denotes the coolant.

The nanofluid density is calculated as follow:

$$\rho_{nf} = \varphi \rho_{np} + (1-\varphi)\rho_{bf} \qquad (6)$$

Where $\varphi$ is the volume fraction of nanoparticles, subscript $nf$ denote nanofluids, $np$ denote nanoparticles, and $bf$ denote base fluid.

The specific heat capacity of the nanofluid is as follows:

$$c_{nf} = \frac{[\varphi \rho_{np} c_{np} + (1-\varphi)\rho_{bf} c_{bf}]}{\rho_{nf}} \tag{7}$$

The thermal conductivity of the nanofluid is as follows:

$$k_{nf} = \frac{k_{np} + 2k_{bf} - 2\varphi(k_{bf} - k_{np})}{k_{np} + 2k_{bf} + \varphi(k_{bf} - k_{np})} \times k_{bf} \tag{8}$$

The variation of inlet flow rate in the enhanced cooling method is divided into three stages as shown in Fig. 3. The initial stage HBTMS is cooled by nanofluid at a constant flow rate $v_m$. The intensive stage is after the PCM starts liquefaction ($t=t_E$). In this stage a step-response coupled Gaussian pulse inlet flow rate function (COMSOL function customization) was developed, which is able to increase the system's ability to continuously dissipate heat. Also, in order to minimize the pumping power consumption of the HBTMS, the plateau is entered when the average battery surface temperature drops to 40°C.

The expression of the enhanced cooling function is as follows:

$$f_E = \begin{cases} v_m & (0 < t < t_E) \\ an_{(t)} \cdot step_{(t)} & (t_E \leq t) \\ v_m & (T \leq T_E) \end{cases} \tag{9}$$

where the subscript E is the enhanced cooling method, $f$ is the flow rate function, $an_{(t)}$ is the Gaussian pulse resolving equation, and $step_{(t)}$ is the step response function. The GS pulse has a time wavelength of 6s and a peak value of 0.1g/s. And the research parameters are $v_m = 0.6g/s$, $t_E = 250s$, $T_E = 40°C$。

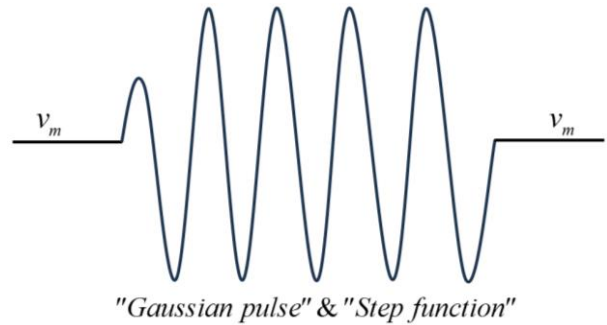

**Fig. 3.** Step pulse flow function of the enhanced cooling method

**2.4 PCM/Aluminum Foam Cooling Method**

The transient phase change process of PCM is simulated using the enthalpy-porosity method, which is commonly used in many literature studies [38,44]. In the enthalpy-porosity approach, the solid-liquid mushy zone is treated as a porous medium with a porosity equaling the volume fraction of the PCM in the liquid phase. As the PCM transforms from solid to liquid, the porosity changes from 0 to 1 accordingly. Moreover, it is reasonable to assume a negligible nature convection effect in the PCM/aluminum foam composite due to the small characteristic length of the porous space in the aluminum foam [23]. The porosity of aluminum foam in this study is 0.95 and the thermal conductivity is 202.4 $W/m \cdot K$. The detailed thermophysical parameters are shown in Table 1.

The energy conservation equation for the PCM/aluminum foam composite is as follows:

$$\rho_{PCM} \frac{\partial H}{\partial t} = \nabla(k_{PCM/Al} \nabla T) \tag{10}$$

Where $H$ is the total enthalpy of PCM, and $k_{PCM-Al}$ is the effective thermal conductivity of PCM/aluminum foam composite.

**Table 1** Thermophysical parameters of different materials

| Material | $\rho(kg \cdot m^{-3})$ | $c(J \cdot kg^{-1} \cdot K^{-1})$ | $k(W \cdot m^{-1} \cdot K^{-1})$ | $\mu(Pa \cdot s)$ |
| --- | --- | --- | --- | --- |
| Battery | 2500 | 1108 | 28 | — |
| Aluminum | 2719 | 871 | 202.4 | — |
| Nanofluid | 1002.91 | 4157.79 | 0.6099 | 0.0009 |
| Paraffin wax RT35 | 770 | 2460 | 5.622(solidity) 0.1505(liquid) | — |

The total enthalpy $H$ of PCM is the sum of sensible and latent enthalpies, which can be calculated as follow:

$$H = \begin{cases} \int_{T_{ini}}^{T} c_{PCM} dT & (T \leq T_S) \\ \int_{T_{ini}}^{T} c_{PCM} dT + \xi \Delta H & (T_S < T \leq T_l) \\ \int_{T_{ini}}^{T_S} c_{PCM} dT + \xi \Delta H + \int_{T_l}^{T} c_{PCM} dT & (T > T_l) \end{cases} \quad (11)$$

The melt fraction $\xi$ is calculated by

$$\xi = \begin{cases} 0 & (T \leq T_S) \\ \dfrac{T - T_S}{T_l - T_S} & (T_S < T \leq T_l) \\ 1 & (T > T_l) \end{cases} \quad (12)$$

**2.5 Boundary conditions and validation**

The relationship between the sub-models of the lithium battery hybrid thermal management system is shown in Fig. 4. The heat source of the system is the EV battery cell. The heat dissipation technology utilizes a hybrid approach of nanofluid cooling coupled with PCM cooling. When the battery pack is "thermally overloaded", a novel pulse flow function (enhanced cooling) can effectively improve the overall heat dissipation capability of the system.

Based on the above methodology, it is reasonable to make the following assumptions:

(1) The internal contact thermal resistance of the Li-ion HBTMS is neglected.

(2) The thermophysical parameters of the battery, coolant, and PCM/aluminum foam do not vary with temperature.

(3) Volume expansion of PCM during melting is ignored and natural convection after melting is not considered.

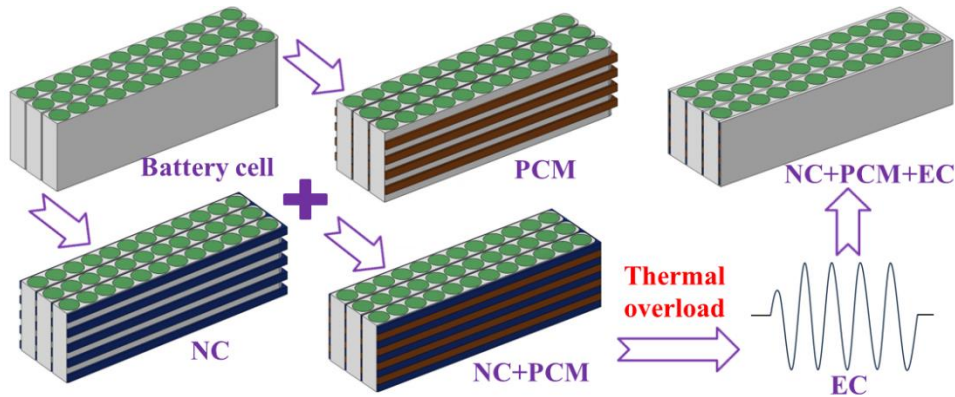

**Fig. 4.** Submodel schematic diagram of hybrid thermal management system for Li-ion batteries

In this paper, the research model and the theoretical governing equations are solved by the commercial software COMSOL Multiphysics 6.0 based on the iterative solution of the flexible generalized minimum residual method (FGMRE). The transient relative tolerance is $10^{-5}$ and the time step is 1s. The model is built parametrically. Solid and fluid heat transfer modules were used. Adding a multi-physics field interface for non-isothermal flows. The 18650 cylindrical battery is set as the heat source and the global convective heat flux of the system is added with a heat transfer coefficient of $5\ Wm^{-2}K^{-1}$. The initial and ambient temperatures of the system were 25°C and the inlet mass flow rate was $v_m = 0.6 g/s$. Setting the liquid outlet as a pressure outlet $p_0 = 0\ Pa$. The maximum Reynolds number of coolants in the inlet channel is less than 2000, so a laminar flow model is used. The phase transition region is modeled as a porous medium in thermal equilibrium. The porous matrix is metallic aluminum. Paraffin RT35 was chosen as the phase change material, which has a large latent heat value, and the ideal phase change temperature matches the range of lithium battery operating ambient temperature. PCM is a liquid fluid after endothermic melting. The initial phase transition temperature is 35°C, and the transition interval is 2°C.

**2.6 Model validation**

It is important to analyze the level of numerical error involved in CFD research through comparative validation to ensure the reliability of the calculation results. It is generally believed that there are three kinds of numerical errors in the calculation of unsteady flow: discrete error, iterative error, and rounding error. The main contribution of numerical error is usually discretization error. Fig. 5 shows the variation curve of the average surface temperature of 3C discharge batteries under different grid quantities. The influence of five precision grids $(42,65,98,164,245) \times 10^4$ on the average temperature of HBTMS was analyzed at the beginning of phase transformation, after phase transformation, and at the end of discharge. It can be found that with the gradual increase of the number of grids, the range of average temperature change gradually decreases. When the number of grids reaches 1.64 million (Fig. 6), the average temperature of the system changes slightly. Therefore, the fourth precision mesh is selected to ensure that the influence of other calculation errors is negligible relative to the total error.

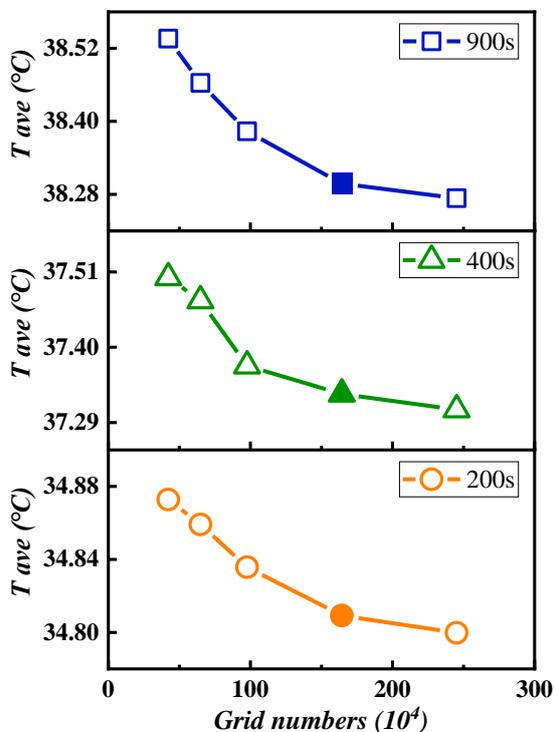
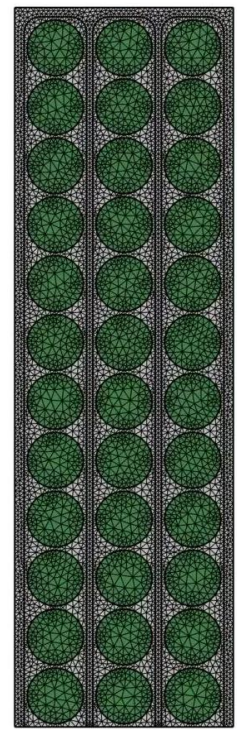

**Fig. 5.** Grid validation of HBTMS          **Fig. 6.** Grid model of $(GN = 164 \times 10^4)$

In order to verify the reliability of the HBTMS heat dissipation model established in this study, it is necessary to compare the numerical model of Li-ion batteries with the heat production of real Li-ion ternary batteries at different discharge rates. When the battery is discharged at a low multiplication rate of $0.1C, 0.2C, 0.3C$, the maximum temperature on the surface of the battery obtained from the simulation is compared with the results of the vehicle test. A pure electric vehicle equipped with Ningde Times Li-ion ternary batteries was used for the whole vehicle discharge test (Fig.7). The vehicle had a theoretical range of 300 km, a rated capacitance of 87 Ah, and a rated voltage of 358 V. The test was started at 100% SOC of the battery pack. Driving at a uniform speed for 15 min, the power battery pack is regarded as discharged. Due to the unavoidable differences between the assembled structure of the vehicle-mounted Li-ion ternary power battery and the theoretical model, as well as the actual discharging process of the battery, there is a loss of power consumption (about 70% of the rated capacity). Combined with the test road and the conditions of the vehicle, the discharge multiplier calculation formula is deduced as:

$$(0.1, 0.2, 0.3)C = \begin{cases} \dfrac{I_{0.1C}}{I_r} \approx \dfrac{v_{0.1C}}{\alpha v_r} & (v_{0.1C} = 21\ km/h) \\ \dfrac{I_{0.2C}}{I_r} \approx \dfrac{v_{0.2C}}{\alpha v_r} & (v_{0.2C} = 42\ km/h) \\ \dfrac{I_{0.3C}}{I_r} \approx \dfrac{v_{0.3C}}{\alpha v_r} & (v_{0.3C} = 63\ km/h) \end{cases} \quad (13)$$

The subscript r is the rated value, is the battery pack current, is the traveling speed, and is the discount ratio.

A thermometer was used to record the surface temperature of the battery pack of the test vehicle, which has its own liquid cooling system (average heat dissipation efficiency of about 0.8), so that the actual temperature was:

$$T_{rea} = \dfrac{T_{te}}{\eta} \quad (\eta = 0.8) \quad (14)$$

$T_{rea}, T_{ve}$ are the actual and measured temperatures of the battery pack surface, respectively. $\eta$ is the thermal efficiency.

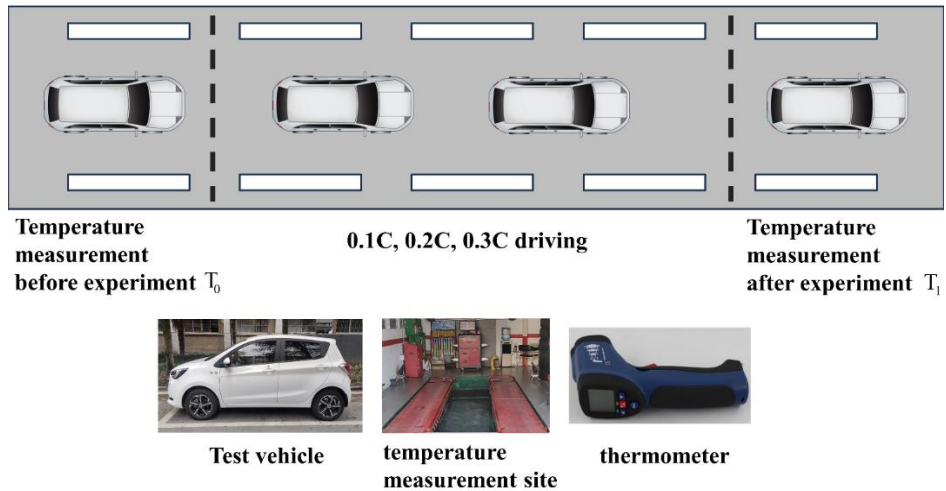

Fig. 7 0.1C, 0.2C, 0.3C whole vehicle discharge test

In order to ensure the accuracy of the test as well as to avoid the distortion of the test data. The test time was from noon to afternoon on a day in the fall (ambient temperature 20~25°C), and the initial temperature of the battery pack. The experiment was conducted five times at three vehicle speeds, totaling 15 tests, and the average value of the highest temperature of the battery pack was taken. The simulation calculation results were compared with the actual maximum temperature of the battery of the whole vehicle test. As shown in Fig. 8, the higher the speed of the pure electric vehicle under the same conditions, the higher the discharge rate of the battery pack, and the higher the maximum temperature of the battery surface. The comparison results in Fig. 8 show that the theoretical model of battery heat production constructed by the simulation is consistent with the trend of the maximum temperature on the battery surface as shown in the vehicle test results. The maximum error of the two methods is 14.3%, which meets the acceptable range of practical error. Overall, the possible factors for this error are fluctuations in the outdoor ambient temperature compared to the theoretically calculated constant temperature; unstable test road conditions (the road contains varying degrees of air humidity and natural wind, etc.); insufficient heat dissipation from the vehicle's liquid-cooling system, and so on.

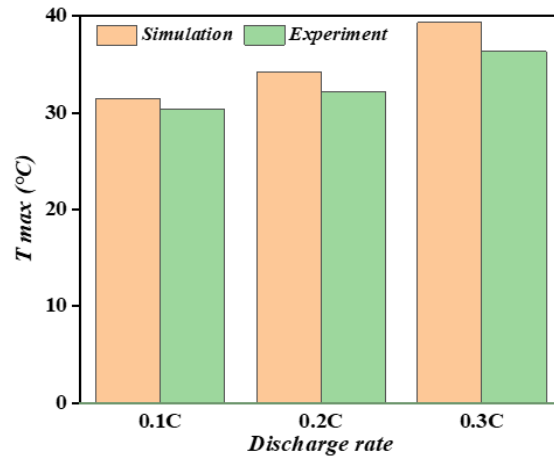

**Fig. 8** Comparison of simulation and test results under 0.1C, 0.2C, 0.3C discharge

When the battery is discharged at $1C, 2C, 3C$ medium to high rates, the maximum temperature on the surface of the battery obtained from the simulation is compared with the experimental results for Qi [45]. Three lithium-ion batteries of the same model were randomly selected in the experiment, with rated capacitance and voltage of 2.6(Ah) and 3.65(V), respectively. The temperature sensing line was placed at three distribution points of the positive electrode, the middle section, and the negative electrode of the single battery, and then tested at $1C, 2C, 3C$ discharge rate in the incubator. The simulation results of different discharge rates are consistent with the experimental results reported by Qi (Fig. 9), and the maximum error is less than 4.8%. Therefore, the simulation results are considered reliable. Combined with the topic of this paper, the thermal performance of BTMS in EV under accelerated state is studied by 3C discharge. Among them, the simulation value is slightly higher than the experimental value. This discrepancy arises because the thermal physical property parameters of lithium-ion batteries and coolants under different discharge rates in the experimental test are affected by temperature, and the convection heat flux coefficient between the system and the environment in the constant temperature box is unstable.

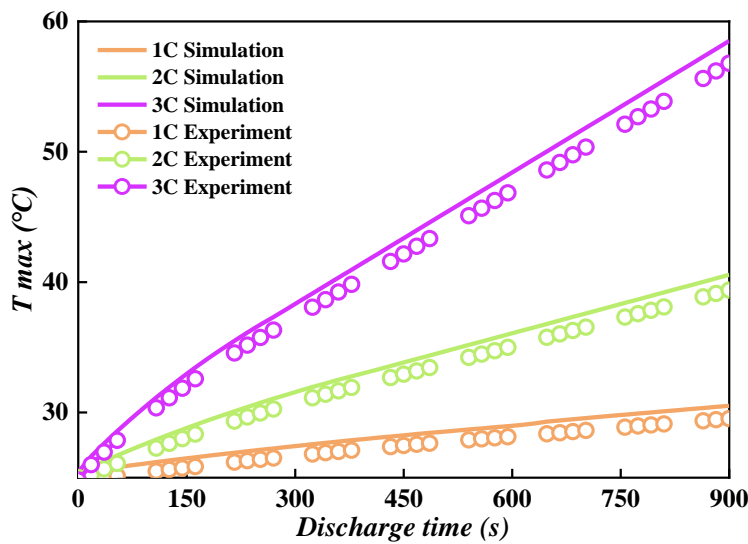

**Fig. 9.** Comparison of simulation results with experimental data for Qi[45]

## 3. Results and discussion
### 3.1 Effect of coolant type on the HBTMS

In various research on battery thermal management, the hybrid cooling method of liquid-cooled coupled phase transition is generally considered to be a cooling method with high heat dissipation efficiency and safety. Because liquid cooling is the dominant heat dissipation mode, it is important to analyze the effect of coolant type on the surface temperature and pumping power consumption of lithium batteries. In addition, with the research of nano-fluids, nano-fluid materials with high thermal conductivity have been obtained, which makes them play a more unique role in the thermal management system of lithium-ion batteries. Therefore, the aqueous base solution of copper oxide, alumina and titanium oxide nanoparticles (0.2% concentration) is used as three research models in this section. Compared with common 50% glycol, kerosene and water, the inlet flow rate is 6g/s, and the specific parameters are shown in Table 2.

The maximum temperature of the battery surface and the change curve of pumping power consumption under the action of six coolants are shown in Fig. 10. Among them, the pumping power consumption of lithium-ion HBTMS is affected by discharge time, coolant pressure drop and volume flow rate, and the calculation expression is as follows:

$$P = \Delta p \cdot V \cdot t \tag{12}$$

It can be observed that when kerosene and ethanol are used as the coolant of HBTMS, their thermal conductivity is relatively low, and the maximum surface temperature of the battery reaches 55.94 °C and 44.53°C after discharge. And because the dynamic viscosity of kerosene and ethanol is larger, the fluid pressure drop in the channel increases, which significantly increases the pumping power consumption. When water and three nanofluids are used as coolants, the maximum surface temperature of the battery is maintained near 40.5°C. However, among these four coolants, the alumina nanofluid showed better heat dissipation performance, the lowest discharge end temperature was 40.32°C, and the increase of pumping power consumption was not obvious.

As a passive cooling method, PCM begins to absorb heat when the surface temperature of the battery reaches the initial value of phase transition. Fig. 11 shows the liquid phase volume fraction of PCM under the action of different coolants. When the discharge time is 200s (the initial phase transition), the liquid phase volume fraction of PCM cooled by kerosene and ethanol has exceeded 0.8. Due to the poor convective heat transfer capacity of these two coolants, the surface temperature of the battery continues to rise, so the PCM can only continue to absorb heat and melt, making the liquid phase volume fraction high. At the same stage (t=200s), the PCM liquid phase volume fraction of the interaction between the nanofluid and water is maintained at about 0.5. When the discharge time is 400s (late phase transition), the heat released by the lithium battery continues to increase with the increase of time. HBTMS In order to further absorb heat and reduce the surface temperature of the battery, the PCM acted by six coolants at this stage continuously consumes latent heat and melts, so the liquid phase volume fraction reaches 0.9.

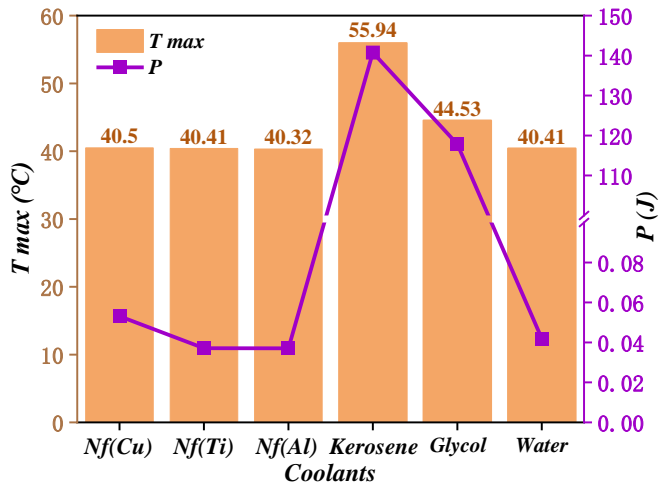
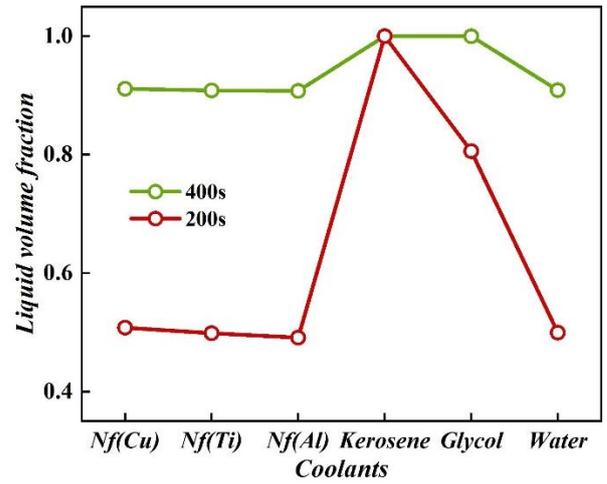

Fig. 10. Maximum battery temperature and system pumping power consumption for different coolants

Fig. 11. Liquid phase volume fraction of PCM for different coolants

Table 2 Thermophysical properties of different coolants [32,45]

| Coolants | $\rho(kg \cdot m^{-3})$ | $c(J \cdot kg^{-1} \cdot K^{-1})$ | $k(W \cdot m^{-1} \cdot K^{-1})$ | $\mu(Pa \cdot s)$ |
|---|---|---|---|---|
| Nf(Cu) | 1008.3 | 4135.7 | 0.6098 | 0.00131 |
| Nf(Ti) | 1003.3 | 4154.9 | 0.6090 | 0.0009 |
| Nf(Al) | 1002.91 | 4157.79 | 0.6099 | 0.0009 |
| Kerosene | 785.7 | 2100 | 0.15 | 2.21 |
| Glycol | 1071.1 | 3300 | 0.38 | 3.39 |
| Water | 998.2 | 4182 | 0.6 | 0.001 |

### 3.2 Effect of cooling direction on the HBTMS

The change of cooling direction makes the convective heat transfer ability of the coolant to the channel wall different, which indirectly affects the energy storage efficiency of HBTMS and the service life of lithium-ion batteries. As shown in Fig. 12, a U-shaped composite channel (composite number n = 3) is designed, consisting of two pairs of entrances and exits. The U-shaped channel matrix is in the wire frame. When applying the combination number of different power batteries, the corresponding matching requirements can be met by changing the channel size and the composite number. In this section, $Nf(Al)$ coolant with inlet flow rate of 6g/s and discharge time of 15min was used to research six different liquid cooling directions.

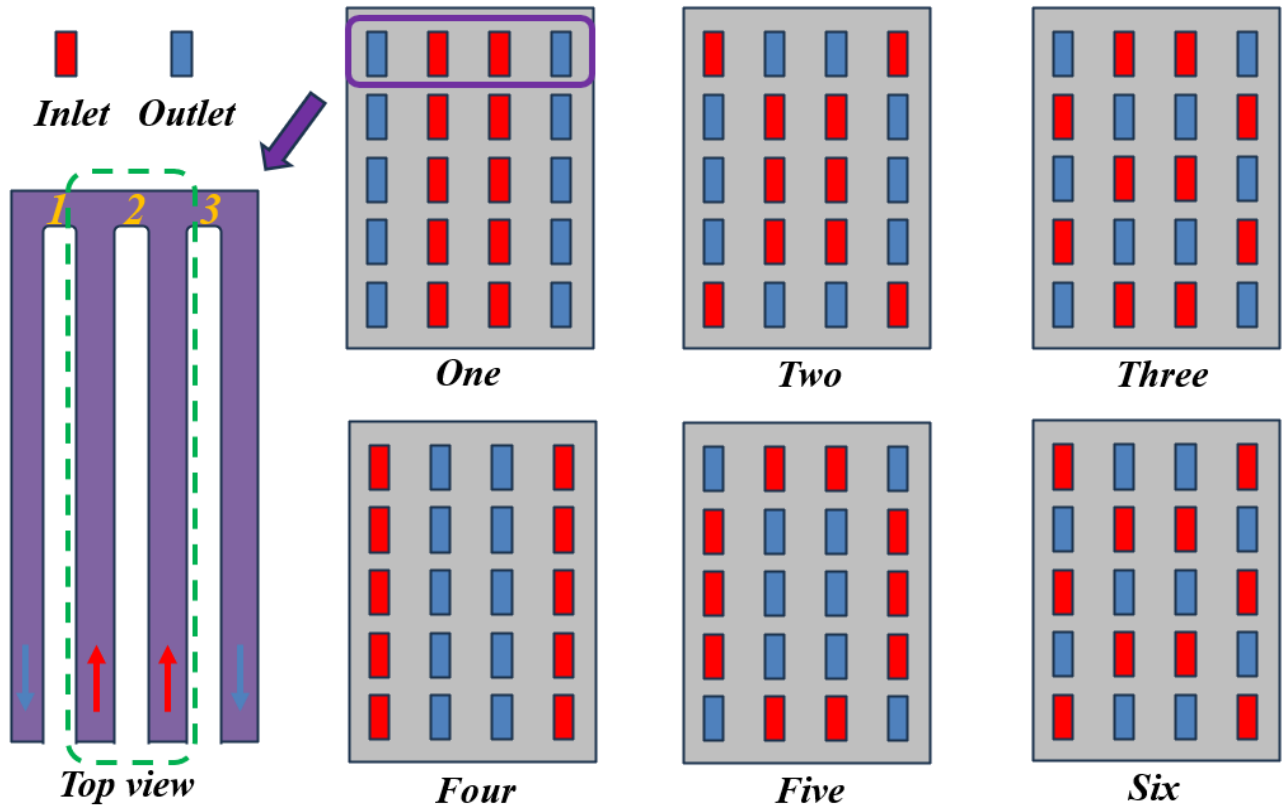

**Fig. 12.** Different cooling directions of U-shaped composite channels

The maximum temperature and pumping power consumption of HBTMS under six cooling directions are shown in Fig. 13. It can be seen that when the coolant is cooled in the fourth direction, the minimum battery surface temperature is 40.34°C. It is worth noting that the system exhibits near-uniform pumping power consumption for all six cooling directions. This is mainly due to the fact that for the same discharge time and inlet flow rate, the cooling direction does not greatly affect the pressure drop of the fluid.

Fig. 14 shows the liquid phase volume fraction of PCM under different cooling directions. For the same discharge time, the surface temperature of the system cooled in the fourth direction is lower. And the liquid phase volume fraction of PCM is smaller than that in other cooling directions, which makes the residual latent heat of PCM higher. Therefore, the fourth cooling direction can provide HBTMS with greater heat dissipation capacity.

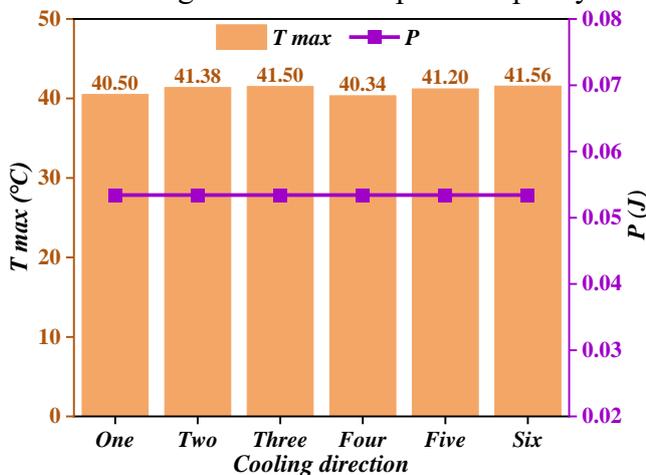

**Fig. 13.** Maximum battery temperature and system pumping power consumption for different cooling directions

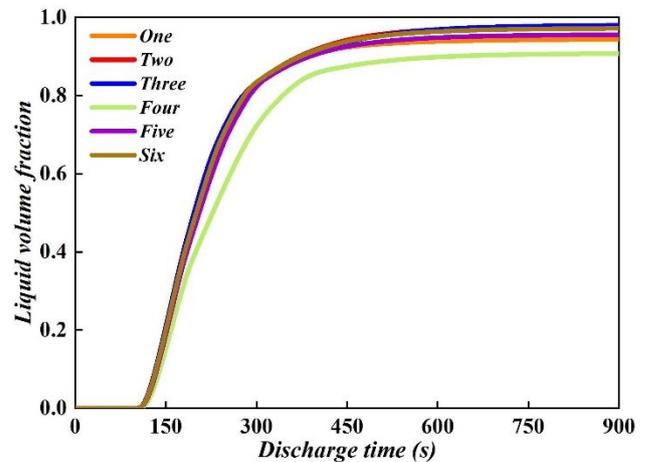

**Fig. 14.** Liquid phase volume fraction of PCM for different cooling directions

## 3.3 Effect of channel height on the HBTMS

This section explores the effect of channel height $D$ on HBTMS. Using $Nf(Al)$ coolant, the fourth cooling direction, an inlet flow rate of 6 g/s and a discharge time of 15 min. In HBTMS, the variation in channel height is simultaneously linked to the fill volume of the PCM, which further affects the overall thermal efficiency of this hybrid system. The variation curves of the channel height on the maximum battery surface temperature and pumping power consumption are shown in Fig. 15. It can be observed that as the height of the channel increases, the maximum battery surface temperature decreases and then increases, but the pumping power consumption gradually decreases. When the channel height D=7mm, the battery surface temperature obtains the lowest value of 40.5°C compared to the maximum channel height of 10mm, and the power consumption increases by only 0.019J. Therefore, in terms of average gain, the overall efficiency of the Li-ion HBTMS is higher at a channel height of 7 mm.

Fig. 16 shows the liquid phase volume fraction of PCM at different channel heights. When the discharge time is less than 380 seconds, a smaller channel height leads to slower liquefaction of the PCM and a greater ability to sustain heat dissipation. Conversely, when the discharge time exceeds 380 seconds, a greater channel height results in faster liquefaction of the PCM and a lower sustained heat dissipation capability. In fact, the nature of this reversal phenomenon is that HBTMS has reached the temperature of PCM near 380s. At this time, the melting degree of PCM is high, and the liquid phase volume fraction reaches more than 0.9. Therefore, after that moment, the main way of HBTMS to maintain heat dissipation is nanofluid liquid cooling, so the higher the fluid channel, the stronger the cooling capacity of the system.

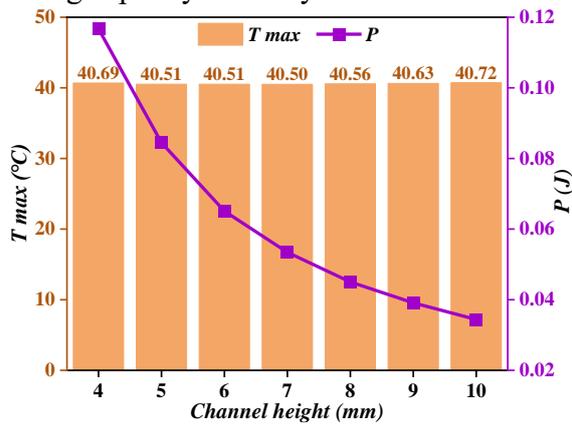 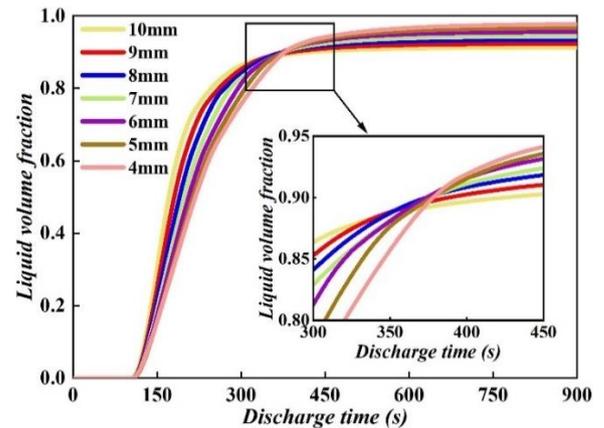

**Fig. 15.** Maximum battery temperature and system pumping power consumption for different channel heights

**Fig. 16.** Liquid phase volume fraction of PCM for different channel heights

## 3.4 Effect of inlet flow rate on the HBTMS

It is well known that the magnitude of the inlet flow rate of a fluid has been used as a more important influencing factor in all theoretical scientific studies that include fluid flow. In this section, the power consumption as well as the thermal performance of the HBTMS is explored by varying the inlet flow rate. $Nf(Al)$ coolant is used, fourth cooling direction, channel height D=7mm, discharge time 15min.

The effects of different constant inlet flow rates on the maximum battery temperature and system pumping power consumption are shown in Fig. 17. As the inlet flow rate increases, the maximum surface temperature of the lithium-ion battery gradually decreases, and the pumping power consumption gradually increases. where the maximum temperature decreases slowly after the flow rate reaches 0.6 g/s, while the pumping power consumption rises faster. The results show that for every degree decrease in the surface temperature of the lithium-ion battery, the pumping power consumption of the system resulting from the increased flow rate rises by an average of about 37.7%. combined with the liquid phase volume fraction of PCM (Fig. 18). At flow rates lower than 0.6 g/s, the liquid phase volume fraction of PCM is higher and the overall heat absorption capacity is reduced. On the contrary, the liquid phase volume fraction of PCM is lower and more heat absorption capacity exists under passive cooling. However, more power consumption is paid for in this condition.

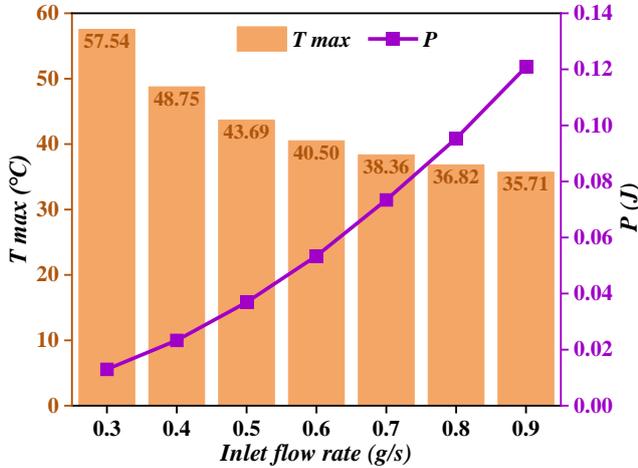

**Fig. 17.** Maximum battery temperature and system pumping power consumption for different inlet flow rate

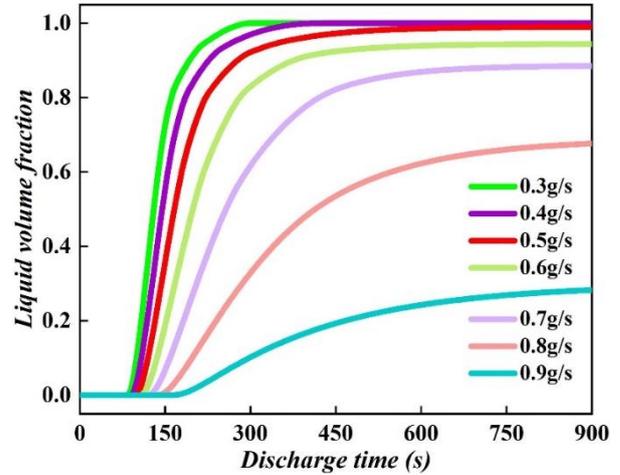

**Fig. 18.** Liquid phase volume fraction of PCM for different inlet flow rate

### 3.5 Effect of cooling method on the HBTMS

From the study of constant inlet flow rate in the previous section (3.4), it is found that the increase of coolant flow can significantly reduce the surface temperature of the battery and improve the heat dissipation capability of the HBTMS. However, it inevitably increases the power consumption of the system. For this reason, it has been thought to dissipate heat by varying the flow rate of cooling [46,47]. In this section, a controlled flow rate function cooling method is discussed. The method aims to regulate the coolant flow rate through "step pulses". This is because the varying flow rate breaks through the thermal boundary layer of convective heat transfer and improves the system's ability to dissipate heat. Using $Nf(Al)$ coolant, fourth cooling direction, channel height D=7mm, discharge time $t_{dis} = 15min$.

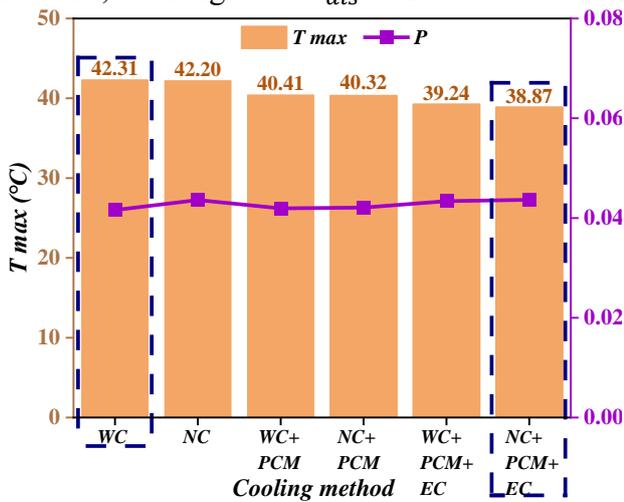

**Fig. 19.** Maximum battery temperature and system pumping power consumption for different

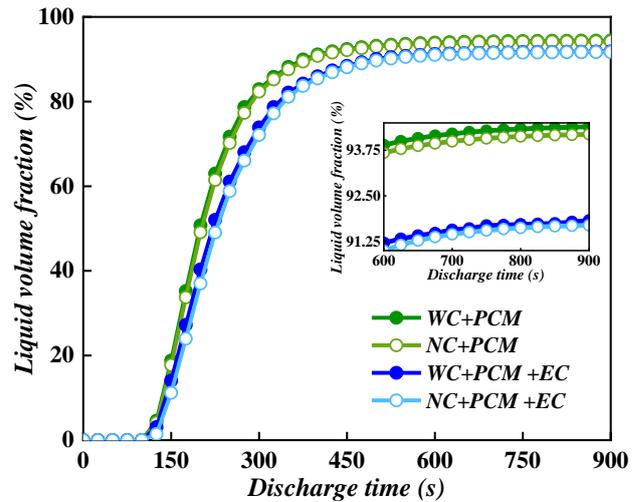

**Fig. 20.** Liquid phase volume fraction of PCM for different cooling methods

The maximum surface temperature of the system, the pumping power consumption, and the liquid phase volume fraction of the PCM for different cooling methods are shown in Fig. 19 and Fig. 20. As can be seen from the figure, the maximum temperature of the battery surface is reduced by the method of mixing liquid cooling and phase change, while the system power consumption increases. Due to the limited space of the battery module in practical engineering applications, it is difficult to change the filling volume of the liquid channel and PCM in the simple hybrid cooling method, at which time the system heat absorption and energy storage rate is constrained. After the battery surface temperature is lowered to a certain threshold, the hybrid cooling method cannot achieve further results. For this reason, an enhanced cooling method is introduced at the beginning of the phase transition. Compared to the HBTMS with constant inlet flow rate, the enhanced method guarantees the continuous heat absorption capacity of the PCM while breaking the heat dissipation temperature threshold. The results show that the maximum temperature on the surface of Li-HBTMS is reduced to 38.87°C on average under the novel enhanced method of NC+PCM+EC. Compared with the normal WC method, the battery temperature is reduced by 3.44°C, and the pumping power consumption of system is increased by only 5%. Based on the research results, the heat dissipation performance of different cooling schemes is converted into the number of battery charges. The new NC+PCM+EC hybrid cooling method can increase the number of battery charges by about 6% to 15% compared with ordinary WC. The battery surface temperature cloud plots and PCM liquid phase fraction cloud plots at different discharge moments under the enhanced cooling method are shown in Fig. 21 and Fig. 22.

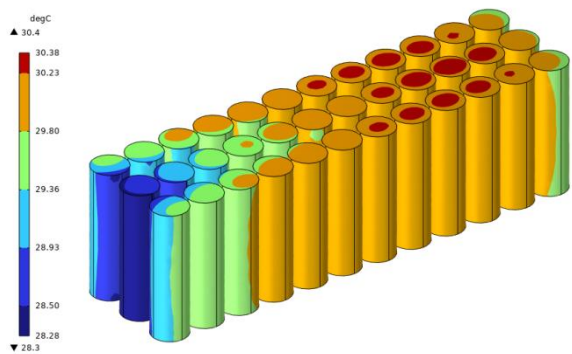
(a) t=50s

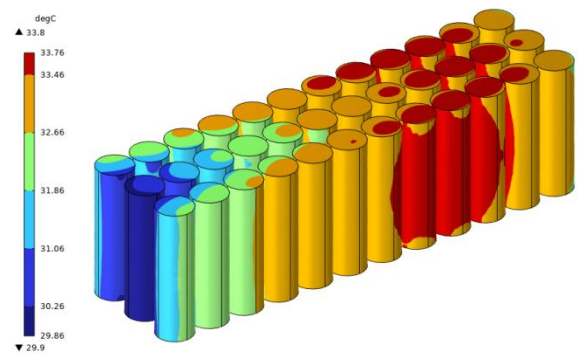
(b) t=100s

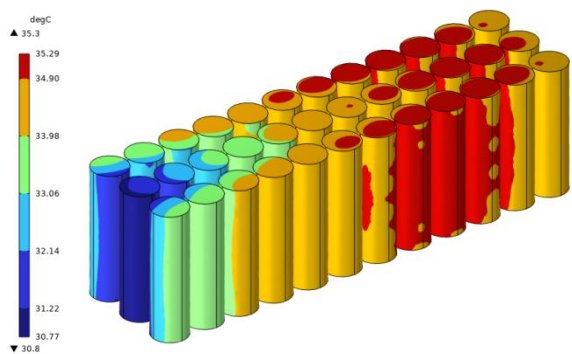
(c) t=150s

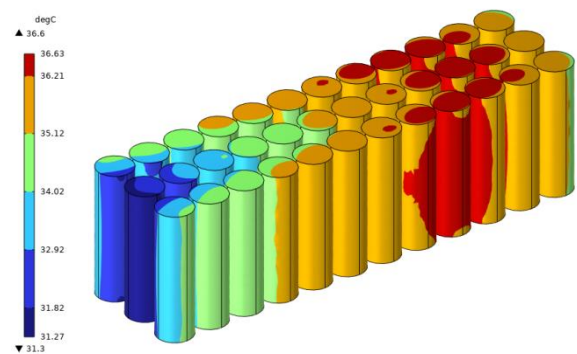
(d) t=250s

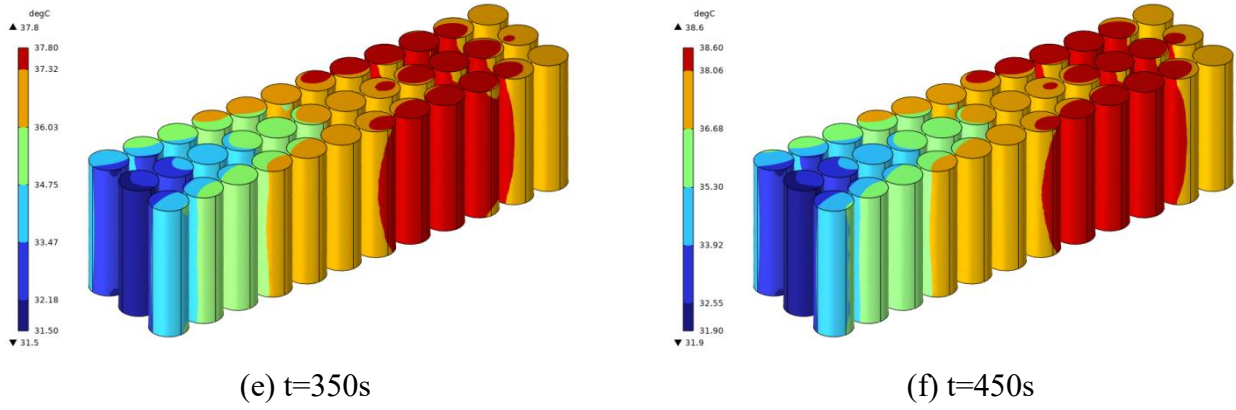

(e) t=350s    (f) t=450s

**Fig. 21.** Maximum surface temperature of the battery at different moments

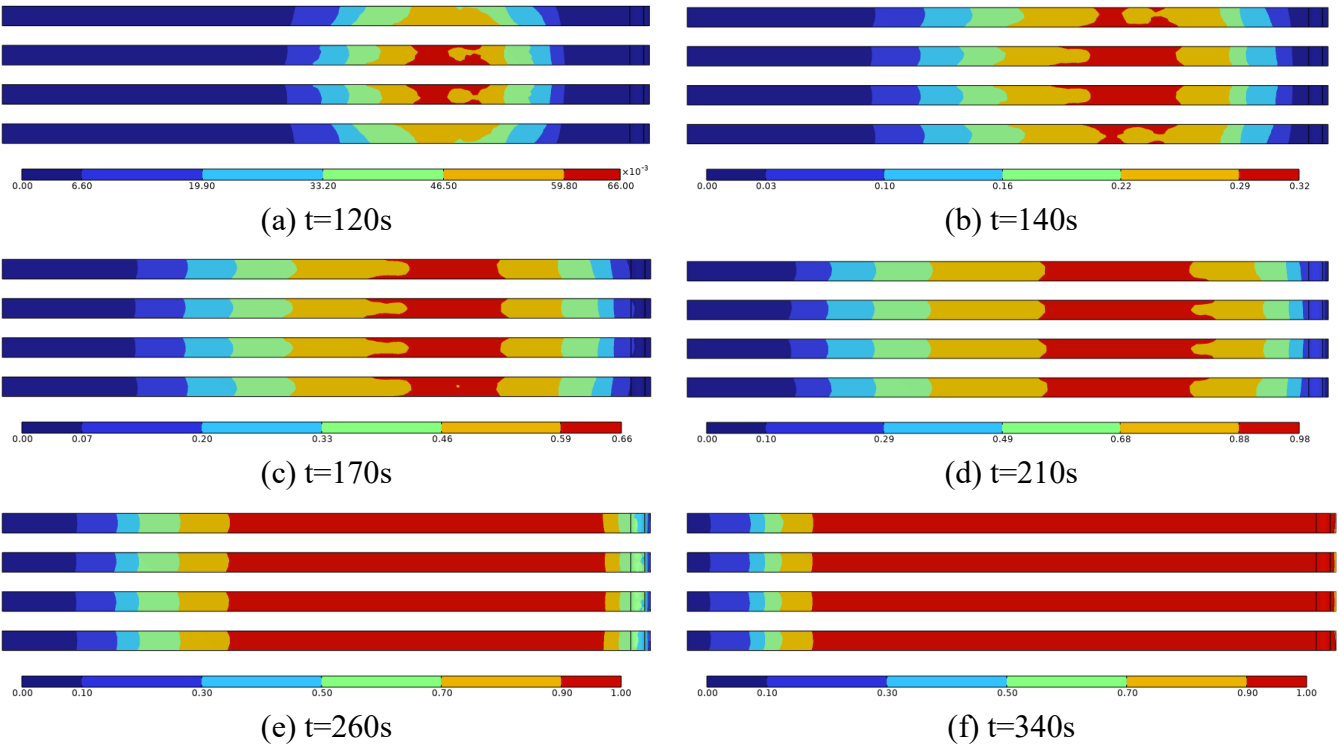

(a) t=120s    (b) t=140s

(c) t=170s    (d) t=210s

(e) t=260s    (f) t=340s

**Fig.22.** Liquid phase volume fraction of PCM at different moments

## 4. Conclusions

Aiming at the thermal management problem generated by high charging/discharging rates of Li-ion batteries, a compact HBTMS integrating U-shaped composite cooling channels with PCM/aluminum foam is proposed. An experimentally validated thermal-fluid dynamics model is developed to comprehensively investigate the effects of different working conditions on the power consumption and thermal performance of the compact HBTMS. The main conclusions are as follows:

1. When using HBTMS with different coolants, *Nf(Al)* has the highest heat dissipation capacity for lithium batteries, with a temperature of 40.32°C at the end of discharge. On the contrary, the overall heat dissipation performance of the system is poor when Kerosene is used.
2. Comparing the six different cooling directions, HBTMS has the best overall heat dissipation performance when the U-shaped composite cooling channel has the two outer sides as the inlet direction.
3. The size of the channel height in HBTMS has an impact on the pumping power consumption

of the system. The higher the channel height, the lower the pumping power consumption of the system, and the better the demonstrated continuous cooling capacity. When the channel height D=7mm, the battery surface temperature obtains the lowest value of 40.5°C compared to the maximum channel height of 10mm, and the power consumption increases by only 0.019J. Therefore, in terms of average gain, the overall efficiency of the Li-ion HBTMS is higher at a channel height of 7 mm.
4. Constant inlet flow rate has a significant effect on the cooling capacity as well as the pumping power consumption of the HBTMS, which increases or decreases inversely with flow rate. Users need to consider the system's heat dissipation and power consumption to make the appropriate decision.
5. A step pulse flow rate function for enhanced cooling was developed. When the new hybrid cooling method of NC+PCM+EC is used, the overall heat dissipation capability of the HBTMS is improved. The results show that the maximum temperature on the surface of Li-HBTMS is reduced to 38.87 °C on average under NC+PCM+EC method. This represents a reduction of 3.44°C compared to the conventional liquid cooling method, which has a maximum surface temperature of 42.31°C. However, the pumping power consumption of the system increases by only 5%.

Future work will investigate in depth the effect of enhanced cooling functions with different varying flow forms (function type, cycle period, flow peak, etc.) on the heat dissipation capacity of the BTMS. And a complete physical experimental study of HBTMS is planned to be carried out when subsequent conditions allow.

**Declaration of Competing Interest**

There are no known competing financial interests or personal relationships that could influence the work reported in this paper.

**Data availability**

Data will be made available on request.

**Acknowledgments**

J.R. Su and L. Chen acknowledge the support of start-up funding at University of Michigan-Dearborn.

# References

[1] G. Zubi, R. Dufo-Lopez, M. Carvalho, G. Pasaoglu, The lithium-ion battery: state of the art and future perspectives, Renew. Sust. Energ. Rev. 89 (2018) 292–308.

[2] B. Jones, R.J.R. Elliott, V. Nguyen-Tien, The EV revolution: the road ahead for critical raw materials demand, Appl. Energy. 280 (2020) 115072.

[3] J.Y. Lin, X.H. Liu, S. Li, C. Zhang, S.C. Yang, A review on recent progress, challenges and perspective of battery thermal management system, Int. J. Heat Mass Transf. 167 (2021) 120834.

[4] P.R. Tete, M.M. Gupta, S.S. Joshi, Developments in battery thermal management systems for electric vehicles: a technical review, J. Energy Storage. 35 (2021) 102255.

[5] Zhao D, Ji C, Teo C, Li S. Performance of small-scale bladeless electromagnetic energy

harvesters driven by water or air, Energy. 74 (2014) 99–108.

[6] Chen Z, Zuo W, Zhou K, Li Q, Huang Y, E J. Multi-objective optimization of proton exchange membrane fuel cells by RSM and NSGA-II, Energy Convers Manag. 277(2023) 116691.

[7] Zhao D, Li S, Yang W, Zhang Z. Numerical investigation of the effect of distributed heat sources on heat-to-sound conversion in a T-shaped thermoacoustic system, Appl. Energy. 144 (2015) 204–13.

[8] Zuo W, Wang Z, E J, Li Q, Cheng Q, Wu Y, Zhou K. Numerical investigations on the performance of a hydrogen-fueled micro planar combustor with tube outlet for thermophotovoltaic applications, Energy. 263 (2023) 125957.

[9] Zuo W, Li D, E J, Xia Y, Li Q, Quan Y, Zhang G. Parametric study of cavity on the performance of a hydrogen-fueled micro planar combustor for thermophotovoltaic applications, Energy. 263 (2023) 126028.

[10] Zuo W, Chen Z, E J, Li Q, Zhang G, Huang Y. Effects of structure parameters of tube outlet on the performance of a hydrogen-fueled micro planar combustor for thermophotovoltaic applications, Energy. 266 (2023) 126434.

[11] Fathy A, Ferahtia S, Rezk H, Yousri D, Abdelkareem MA, Olabi AG. Optimal adaptive fuzzy management strategy for fuel cell-based DC microgrid, Energy. 247 (2022) 123447.

[12] Olabi AG, Wilberforce T, Sayed ET, Abo-Khalil AG, Maghrabie HM, Elsaid K, Abdelkareem MA. Battery energy storage systems and SWOT (strengths, weakness, opportunities, and threats) analysis of batteries in power transmission, Energy. 254 (2022) 123987.

[13] Wilberforce T, Rezk H, Olabi AG, Epelle EI, Abdelkareem MA. Comparative analysis on parametric estimation of a PEM fuel cell using metaheuristics algorithms, Energy. 262 (2023) 125530.

[14] Olabi AG, Abdelkarem MA, Jouhara H. Energy digitalization: main categories, applications, merits and barriers, Energy. 270 (2023) 126899.

[15] Gandoman FH, Jaguemont J, Goutam S, Gopalakrishnan R, Firouz Y, Kalogiannis T, Omar N, Van Mierlo J. Concept of reliability and safety assessment of lithium-ion batteries in electric vehicles: basics, progress, and challenges, Appl. Energy. 251 (2019) 113343.

[16] Gan Y, He L, Liang J, Tan M, Xiong T, Li Y. A numerical study on the performance of a thermal management system for a battery pack with cylindrical cells based on heat pipes, Appl. Therm. Eng. 179 (2022) 115740.

[17] Wu C, Wang Z, Bao Y, Zhao J, Rao Z. Investigation on the performance enhancement of baffled

cold plate based battery thermal management system, J. Energy Storage. 41 (2021) 102882.

[18] Liu C, Wang Y, Chen Z. Degradation model and cycle life prediction for lithium-ion battery used in hybrid energy storage system, Energy. 166 (2019) 796–806.

[19] Feng X, Ren D, He X, Ouyang M. Mitigating thermal runaway of lithium-ion batteries, Joule. 4 (2020) 743–70.

[20] Huang Z, Shen T, Jin K, Sun J, Wang Q. Heating power effect on the thermal runaway characteristics of large-format lithium ion battery with Li(Ni1/3Co1/3Mn1/3)O2 as cathode. Energy. 239 (2022) 121885.

[21] Zuo W, Zhang Y, E J, Huang Y, Li Q, Zhou K, Zhang G. Effects of multi-factors on performance of an improved multi-channel cold plate for thermal management of a prismatic LiFePO4 battery, Energy. 261 (2022) 125384.

[22] Y. Chen, M. Sang, W. Jiang, Y. Wang, Y. Zou, C. Lu, Z. Ma, Fracture predictions based on a coupled chemo-mechanical model with strain gradient plasticity theory for film electrodes of Li-ion batteries, Eng. Fract. Mech. 253 (2021) 107866.

[23] Yang H, Li M, Wang Z, Ma B, A compact and lightweight hybrid liquid cooling system coupling with Z-type cold plates and PCM composite for battery thermal management, Energy. 263 (2023) 126026.

[24] Chen K, Wu WX, Yuan F, Chen L, Wang SF. Cooling efficiency improvement of air-cooled battery thermal management system through designing the flow pattern, Energy. 167 (2019) 781–90.

[25] Chen K, Chen Y, She Y, Song M, Wang S, Chen L. Construction of effective symmetrical air-cooled system for battery thermal management, Appl. Therm. Eng. 166 (2020) 114679.

[26] Liu Y, Zhang J. Self-adapting J-type air-based battery thermal management system via model predictive control, Appl. Energy. 263 (2020) 114640.

[27] Rao ZH, Qian Z, Kuang Y, Li YM. Thermal performance of liquid cooling based thermal management system for cylindrical lithium-ion battery module with variable contact surface, Appl. Therm. Eng. 123 (2017) 1514–22.

[28] Park S, Jang DS, Lee D, Hong SH, Kim Y. Simulation on cooling performance characteristics of a refrigerant-cooled active thermal management system for lithium ion batteries, Int. J. Heat Mass Transfer. 135 (2019) 131–41.

[29] Wu X, Du J, Guo H, Qi M, Hu F, Shchurov NI. Boundary conditions for Onboard thermal-management system of a battery pack under ultrafast charging, Energy. 243 (2022) 123075.

[30] Luo W, Li H, Chu T, A numerical study of battery thermal management system with square spiral ring-shaped liquid cooling plate, Thermal Science and Engineering Progress. 45 (2023) 102120.

[31] Bao Y, Shao S, Numerical study on ultrathin wide straight flow channel cold plate for Li-ion battery thermal management, J. Energy Storage. 64 (2023) 107263.

[32] Tareq S, Malek A, Abdul Ghani Olabi, Ahmed A, Mohammad A, Experimental and numerical analysis of heat transfer enhancement inside concentric counter flow tube heat exchanger using different nanofluids, International Journal of Thermofluids. 20 (2023) 20100432.

[33] A.D. Tuncer, A. Khanlari, A. Sozen, E.Y. Gürbüz, H.I. Variyenli, Upgrading the performance of shell and helically coiled heat exchangers with new flow path by using TiO2/water and CuO–TiO2/water nanofluids, Int. J. Therm. Sci. 183 (2023) 107831.

[34] W. Ajeeb, R.R.S. Thieleke da Silva, S.M.S. Murshed, Experimental investigation of heat transfer performance of Al2O3 nanofluids in a compact plate heat exchanger, Appl. Therm. Eng. 218 (2023) 119321.

[35] Liu S, Liu Y, Gu H, Tian R, Huang H, Yu T, Experimental study of the cooling performance of γ-Al2O3/heat transfer fluid nanofluid for power batteries, J. Energy Storage. 72 (2023) 108476.

[36] Husam Abdulrasool Hasan, Hussein Togun, Azher M. Abed, Naef A.A. Qasem, Aissa Abderrahmane, Kamel Guedri, Sayed M. Eldin, Numerical investigation on cooling cylindrical lithium-ion-battery by using different types of nanofluids in an innovative cooling system, Case Studies in Thermal Engineering. 49 (2023) 103097.

[37] Jiang L, Zhang HY, Li JW, Xia P. Thermal performance of a cylindrical battery module impregnated with PCM composite based on thermoelectric cooling, Energy. 188 (2019) 116048.

[38] Ping P, Zhang Y, Kong D, Du J. Investigation on battery thermal management system combining phase changed material and liquid cooling considering non-uniform heat generation of battery, J. Energy Storage. 36 (2021) 102448.

[39] Bai F, Chen M, Song W, Yu Q, Li Y, Feng Z, Ding Y. Investigation of thermal management for lithium-ion pouch battery module based on phase change slurry and mini channel cooling plate, Energy. 167 (2019) 561–74.

[40] Zhou Z, Wang D, Peng Y, Li M, Wang B, Cao B, Yang LZ. Experimental study on the thermal management performance of phase change material module for the large format prismatic lithiumion battery, Energy. 238 (2022) 122081.

[41] He LF, Tang XW, Luo QL, Liao YP, Luo XY, Liu JL, Ma L, Dong D, Gan YH, Li Y. Structure optimization of a heat pipe-cooling battery thermal management system based on fuzzy grey

relational analysis, Int. J. Heat Mass Tran. 182 (2022) 121924.

[42] Wu W, Yang X, Zhang G, Chen K, Wang S. Experimental investigation on the thermal performance of heat pipe-assisted phase change material based battery thermal management system, Energy Convers Manag. 138 (2017) 486–92.

[43] Wang Q, Jiang B, Xue QF, Sun HL, Li B, Zou HM, Yan YY. Experimental investigation on EV battery cooling and heating by heat pipes, Appl. Therm. Eng. 88 (2015) 54–60.

[44] Liu Z, Xu G, Xia Y, Tian S, Numerical study of thermal management of pouch lithium-ion battery based on composite liquid-cooled phase change materials with honeycomb structure, J. Energy Storage. 70 (2023) 108001.

[45] Wenjie Qi, Wenqi Huang, Juntian Niu, Feng Chen, Bin Chen, Yong Chen, Thermal management of power battery based on flexible Swiss roll type liquid cooling micro-channel, Appl. Therm. Eng. 219 (2023) 119491.

[46] Yonghao Wang, Tieyu Gao, Liang Zhou, Jianying Gong, Jun Li, A parametric study of a hybrid battery thermal management system that couples PCM with wavy microchannel cold plate, Appl. Therm. Eng. 219 (2023) 119625.

[47] Dexin Li, Wei Zuo, Qingqing Li, Guangde Zhang, Kun Zhou, Jiaqiang E, Effects of pulsating flow on the performance of multi-channel cold plate for thermal management of lithium-ion battery pack, Energy. 273 (2023) 127250.